\documentclass[a4paper,twocolumn,
prl,
amsmath,amssymb,
superscriptaddress,
showpacs]{revtex4-2}

\usepackage[english]{babel}

\usepackage{booktabs}
\usepackage{tabu}
\usepackage{quantikz}

\usepackage{amsmath}
\usepackage{amssymb}
\usepackage{graphicx}
\usepackage{physics}
\usepackage[breakable,skins]{tcolorbox}
\usepackage{ulem}

\usepackage{color}
\definecolor{blue}{rgb}{0,0.2,1}

\definecolor{red}{rgb}{0.9,0,0}

\usepackage[colorinlistoftodos]{todonotes}
\usepackage[colorlinks=true, allcolors=blue]{hyperref}
\usepackage[sort&compress]{natbib}

\newcommand{\inlineheading}[1]{\textbf{#1~-- }\ignorespaces}

\begin{document}

\title{Variational Quantum Dimension Reduction for Recurrent Quantum Models
}

\author{Chufan Lyu}
\affiliation{Institute of Fundamental and Frontier Sciences,
University of Electronic Science and Technology of China, Chengdu 611731, China}
\affiliation{Nanyang Quantum Hub, School of Physical and Mathematical Sciences, Nanyang Technological University, Singapore 637371}

\author{Ximing Wang}
\affiliation{Nanyang Quantum Hub, School of Physical and Mathematical Sciences, Nanyang Technological University, Singapore 637371}

\author{Mile Gu}
\email{mgu@quantumcomplexity.org}
\affiliation{Nanyang Quantum Hub, School of Physical and Mathematical Sciences, Nanyang Technological University, Singapore 637371}

\author{Thomas J.~Elliott}
\affiliation{Department of Physics \& Astronomy, University of Manchester, Manchester M13 9PL, United Kingdom}
\affiliation{Department of Mathematics, University of Manchester, Manchester M13 9PL, United Kingdom}
\affiliation{Centre for Quantum Science and Engineering, University of Manchester, Manchester M13 9PL, United Kingdom}

\author{Chengran Yang}
\email{yangchengran92@gmail.com}
\affiliation{Nanyang Quantum Hub, School of Physical and Mathematical Sciences, Nanyang Technological University, Singapore 637371}
\affiliation{Centre for Quantum Technologies, National University of Singapore, 3 Science Drive 2, Singapore 117543}

\date{\today}

\begin{abstract}
Recurrent quantum models (RQMs) realize sequential quantum processes through repeated application  of a unitary operation on a memory system coupled with a series of output registers. However, such models often rely on unnecessarily large memory spaces, introducing redundancy and limiting scalability. Here, we introduce a \textit{variational quantum dimension reduction} framework that identifies and removes irrelevant memory degrees of freedom while preserving the recurrent dynamics of the target model. Our approach employs two parameterized quantum circuits: a decoupling unitary $V(\theta_1)$ that isolates the essential memory subspace; and a compressed recurrent unitary $\tilde{U}(\theta_2)$ that reconstructs the dynamics in the reduced space. The optimization is guided by a unified cost function combining decoupling fidelity and dynamical accuracy, evaluated using the \textit{Quantum Fidelity Divergence Rate} (QFDR), a metric that quantifies long-term fidelity per time step. Applied to a cyclic random walk model, our framework achieves up to three orders of magnitude smaller QFDR compared to variational matrix product state truncation, while requiring only trajectory samples rather than explicit state reconstructions. This establishes a scalable, data-driven paradigm for learning minimal recurrent quantum architectures, enabling variational circuit optimization and quantum process compression for near-term quantum devices.
\end{abstract}

\maketitle


\inlineheading{Introduction} Recurrent models play a central role in the study of sequential and temporal processes~\cite{connorRecurrentNeuralNetworks1994,Rabiner89,smouse2010stochastic}. By repeatedly applying a fixed transition rule, typically implemented by a neural network or a stochastic update map, these models generate output sequences with rich temporal correlations. Classical examples include recurrent neural networks and hidden Markov models (HMMs), which have been widely deployed in time-series forecasting~\cite{connorRecurrentNeuralNetworks1994,Pederson1997HiddenMA, hewamalageRecurrentNeuralNetworks2021}, signal processing~\cite{galesApplicationHiddenMarkov2008,Graves2013}, natural language processing~\cite{levinson1986continuously,mikolov10_interspeech}, and the modeling of complex dynamical systems~\cite{Fraser2008,Al-ani11,RecurrentNeuralNetwork1998}. Their success highlights the importance of understanding and efficiently representing recurrent structure in both theory and applications.

This paradigm has recently been extended to the quantum domain through recurrent quantum models (RQMs), such as quantum Hidden Markov models (QHMMs)~\cite{monras2010hidden, Monras16_Quantum, cholewaQuantumHiddenMarkov2017,srinivasan18a,adhikary20a,elliott2021memory,markov2025implementationlearningquantumhidden, zonnios2025quantum, riechers2025identifiability,vieira2022temporal} and quantum causal models~\cite{Binder2018,Liu2019}. In contrast to their classical counterparts, which store information in classical states, quantum recurrent models encode memory in quantum states. This quantum encoding can lead to substantial reductions in the effective memory dimension required to reproduce a given process~\cite{elliott2020extreme,elliott2021quantum,wu2023implementing, Fanizza24_Quantum, yangDimensionReductionQuantum2025}. Moreover, quantum recurrent models naturally allow the generation of so-called q-samples, quantum states that coherently superpose all possible trajectories of the process, thereby enabling intrinsically quantum forms of simulation and inference~\cite{schuld2018quantum, blank2021quantum}.

In this paper, we address the following question: given oracular access to an RQM specified by an unknown coupling unitary, can we (efficiently) reverse-engineer its dynamics using a quantum circuit and systematically simplify its implementation? Since the essential physics of the process is encoded in the observable output sequences, the internal memory system may contain redundant degrees of freedom. This raises the possibility of compression -- reducing circuit depth, gate count, or memory size -- without altering the generated process. Such simplification is not always achievable, as some processes inherently require large resources, but identifying when and how compression is possible is crucial for practical deployment. To this end, we develop a variational decoupling algorithm that searches for resource-efficient circuit implementations that faithfully reproduce the sequential dynamics while minimizing memory usage whenever possible. We demonstrate the effectiveness of this approach by applying it to cyclic random walk processes~\cite{adyan1983random,garnerProvablyUnboundedMemory2017}.

\begin{figure}[ht]
    \centering
    \includegraphics[width=0.9\columnwidth]{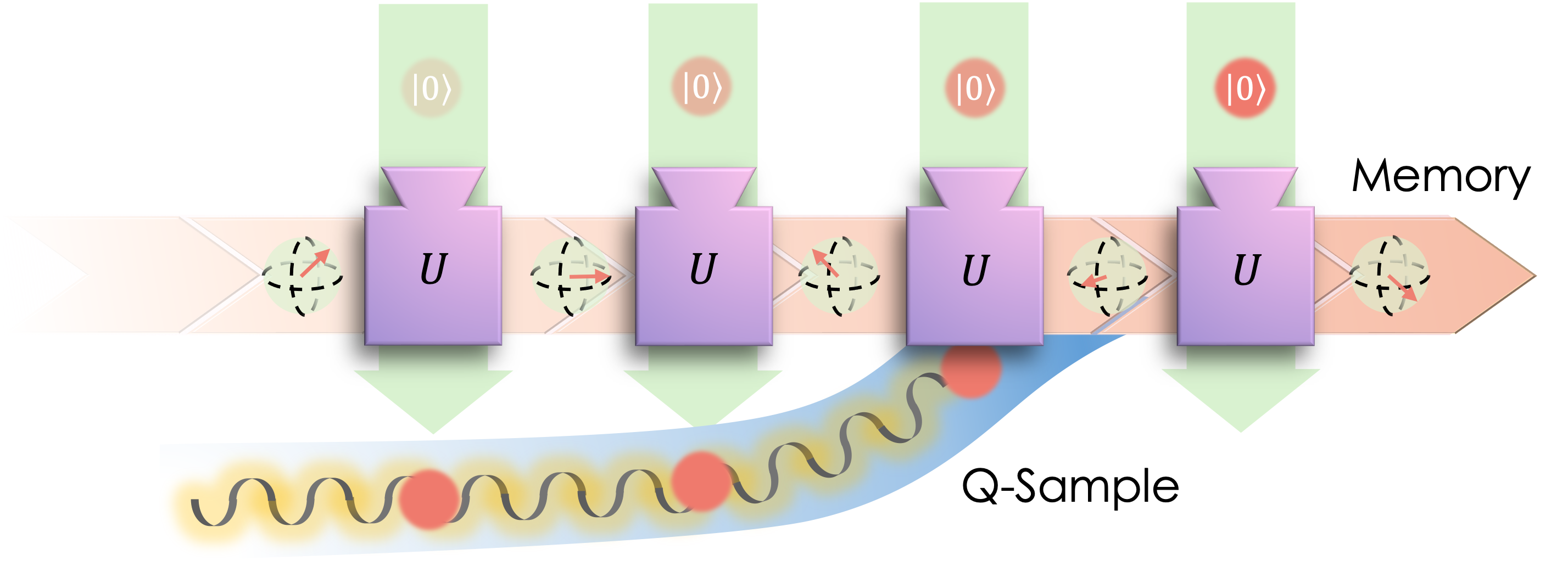}
    \caption[]{Recurrent quantum model: A sequential quantum circuit where a hidden memory state is iteratively updated by repeated unitary blocks $U$ coupling to input qubits initialised in $|0\rangle$ to produce the target data stream (Q-Sample).}
    \label{fig:RQM circuit}
\end{figure}

\inlineheading{Framework} RQMs enable the sequential generation of one-dimensional quantum states, such as 1D quantum many-body states~\cite{Solano2005} or quantum sample states that represent a superposition of all possible trajectories of a stochastic process~\cite{Yang2018}. Analogous to a classical recurrent neural network, an RQM comprises two primary components: a sequence of $L$ output registers and an internal memory system that mediates correlations between these outputs (see~Fig.~\ref{fig:RQM circuit}). At each time step, a fixed unitary operator $U$ acts on the joint state of the current memory $\ket{\sigma_i}$ and a new output register initialized in a vacuum state (e.g., $\ket{0}$):

\begin{equation}
U\ket{\sigma_i}\ket{0} = \sum_x c_{x|i}\ket{\sigma (x,i)} \ket{x},
\end{equation}
where $c_{x|i}$ denotes the transition amplitude and $\ket{\sigma (x,i)}$ represents the updated memory state conditional on the output $\ket{x}$.

After applying the unitary operator $U$ for $L$ time steps, the collection of all $L$ output registers yields an entangled quantum state $\ket{\Psi}$ that represents a coherent superposition of all possible trajectories:
\begin{equation}
    \ket{\Psi} = \sum_{x_{0:L}} c_{x_{0:L}} \ket{\sigma(x_{0:L})} \ket{x_{0:L}},
\end{equation}
where $x_{0:L} = x_0 \dots x_{L-1}$ denotes the output sequence, $\ket{\sigma(x_{0:L})}$ is the corresponding memory state conditioned on the sequence $x_{0:L}$, and we have suppressed the explicit dependence on the initial memory state for ease of notation.
A key advantage of RQMs is that the required circuit depth scales linearly with $L$, in contrast to naive direct state-preparation schemes which typically require resources exponential in $L$.

In the asymptotic limit, the reduced memory state converges to a stationary state $\rho_M$:
\begin{equation}
    \rho_M = \lim_{L\to\infty}\sum_{x_{0:L}} |c_{x_{0:L}}|^2 \ketbra{\sigma(x_{0:L})}{\sigma(x_{0:L})}.
\end{equation}
This stationary state characterizes the asymptotic information stored in the memory and serves as a reference for quantifying memory resources.
Since the memory system propagates information from the past to the future, its dimension directly determines the model's capacity.
Consequently, the implementation cost of an RQM is often quantified via the entropy of $\rho_M$.

To quantify the memory cost, we employ two standard measures~\cite{Liu2019,suenSurveyingStructuralComplexity2022,hoRobustInferenceMemory2020,furrerMinMaxEntropyInfinite2011}: (1) the von Neumann entropy $C_q := S(\rho_M)$, and (2) the max-entropy (or Rényi-0 entropy) $n=\log_2(\mathrm{Dim}[\rho_M])$, corresponding to the physical number of qubits required to embed the memory system.
While the inequality $C_q \leq n$ necessarily holds, in many practical scenarios~\cite{garnerProvablyUnboundedMemory2017,elliottSuperiorMemoryEfficiency2018,elliottMemoryefficientTrackingComplex2019,aghamohammadiExtremeQuantumAdvantage2017} one observes $C_q \ll n$.
This discrepancy indicates significant redundancy in the memory embedding, suggesting that irrelevant degrees of freedom can be eliminated through dimensionality reduction.

\begin{figure}[ht]
    \centering
    \includegraphics[width=0.98\columnwidth]{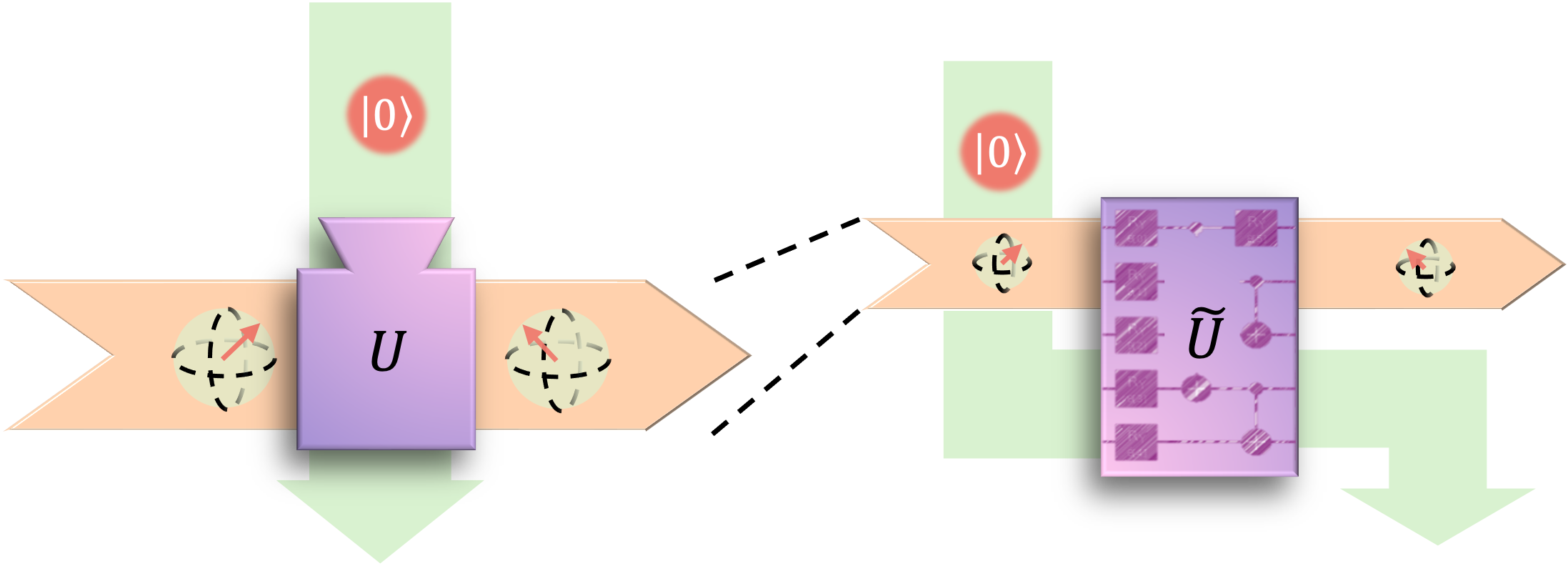}
    \caption[]{
  The objective of this work is to determine an accurate, dimension-reduced approximation of a target RQM governed by a black-box unitary $U$ (left), via a variational quantum circuit realization $\tilde{U}$ (right).
    }
    \label{fig:Objective}
\end{figure}

Our goal is to obtain a full circuit implementation of an RQM with reduced memory dimension $\tilde{n}$ that faithfully reproduces the dynamics generated by a target RQM, whose coupling unitary operator is treated as a black box (see Fig.~\ref{fig:Objective}). The central challenge lies in constructing an approximation that maximizes accuracy under a strict memory budget (i.e., $\tilde{n} < n$ and limited circuit depth), thereby achieving substantial compression without sacrificing dynamical fidelity.

Although the fidelity $\mathcal{F}$ between two quantum states is a natural similarity measure, any small distortion per time step causes the fidelity between the resulting q-sample states to decay exponentially with the sequence length. We therefore use an intensive analogue, the Quantum Fidelity Divergence Rate (QFDR), to quantify the performance of an approximate RQM~\cite{Bram2021,yangDimensionReductionQuantum2025}, as it captures the distortion rate per time step. It is defined as
\begin{equation}\label{eq:qfdr}
    R_F(\ket{\Psi}, \ket{\Psi'}) := -\lim_{L \to \infty} \frac{1}{2L} \log_2 \mathcal{F}(\ket{\Psi_L}, \ket{\Psi'_L}),
\end{equation}
where $\ket{\Psi_L}$ and $\ket{\Psi'_L}$ denote the $L$-step q-samples, and $\mathcal{F}(\cdot,\cdot)$ measures the fidelity between two quantum states. There are many possible fidelity notions for density matrices; in this work, we adopt the cosine similarity between two quantum states, defined as~\cite{wang2008,liang2019}
\begin{equation}\label{eq:cosine_similarity}
    \mathcal{F}(\rho, \sigma) = \frac{\operatorname{Tr}[\rho \sigma]}{\sqrt{\operatorname{Tr}[\rho^2]\operatorname{Tr}[\sigma^2]}}.
\end{equation}
This reduces to the standard fidelity when both states are pure. A lower QFDR indicates closer agreement between the generated sequences over increasing lengths, and hence better performance of the reduced model. In practice, $R_F$ can be efficiently estimated whenever the processes admit compact representations, such as an infinite Matrix Product State (iMPS)~\cite{Bram2021,Yang2020}.

To achieve this reduction in practice, we employ a variational quantum circuit to approximate a low-dimensional update unitary. 
This unitary is realized via a parameterized circuit consisting of single-qubit rotations and entangling gates, serving as a flexible ansatz capable of mimicking the action of the original coupling unitary on the relevant memory subspace. 
The circuit parameters are optimized by minimizing a cost function specifically defined to quantify the discrepancy between the dynamics induced by the original RQM and those produced by the reduced model. 
We perform this optimization using standard hybrid quantum-classical routines~\cite{schuldEvaluatingAnalyticGradients2019,peruzzoVariationalEigenvalueSolver2014,mccleanTheoryVariationalHybrid2016,kandalaHardwareefficientVariationalQuantum2017,crooksGradientsParameterizedQuantum2019,wierichsGeneralParametershiftRules2022}, where parameters are updated iteratively via gradient-based methods (utilizing the parameter-shift rule) or gradient-free optimizers. 
The specific formulation of the cost function and the complete optimization framework are detailed in the next Section.

\begin{figure}[htbp]
    \centering
    \includegraphics[width=\columnwidth]{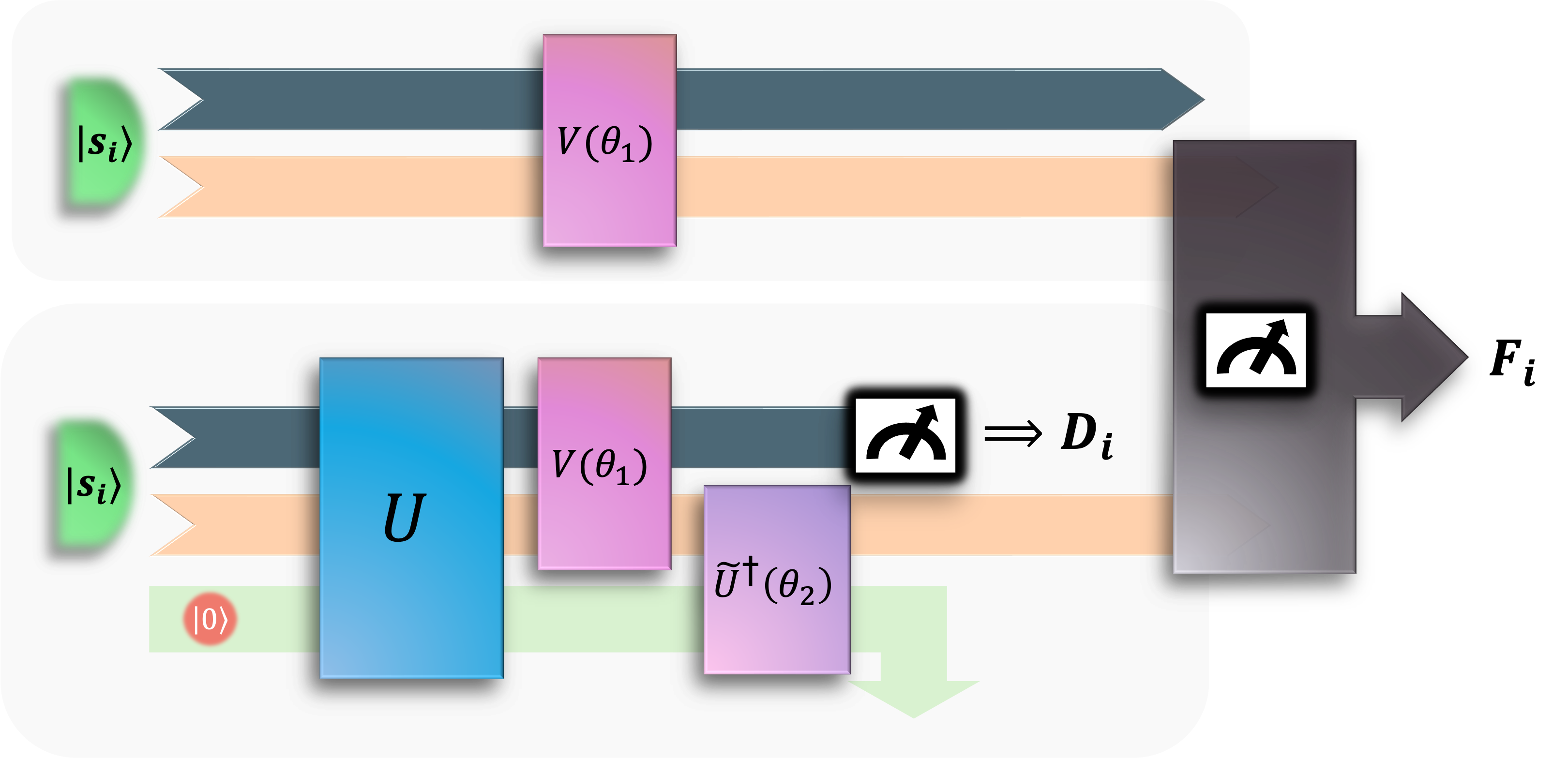}
    \caption[]{Schematic illustration of the circuit constructions used to evaluate the variational cost function. The figure shows, in abstract form, how the original update $U$, the decoupling unitary $V(\boldsymbol{\theta_1})$, and the reduced update $\widetilde{U}(\boldsymbol{\theta_2})$ are combined to generate the quantities entering the cost.} \label{fig:cost}
\end{figure}

\begin{figure}[ht]
    \vspace*{0.5em}
    \begin{tcolorbox}[skin=enhanced]
     {\bf Variational Quantum Circuit Optimization for Memory Reduction in Recurrent Quantum Models}
  
     \vspace*{0.5em}
     {\bf Given:}\\
     \vspace*{0.25em}
     \begin{tabular}{rcl}
        $U$ & -- & (Black box access to) Unitary operator \\ & & for the update function of the target \\ & & recurrent quantum model. \\
        $\{\ket{s_i}\}_{i=1}^K$ & -- & Ensemble of sampled stationary memory \\ & & states of the target model. \\
     \end{tabular}
    
     \vspace*{0.5em}
     {\bf Outputs:}\\
     \vspace*{0.25em}
     \begin{tabular}{rcl}
        $V(\boldsymbol{\theta_1^*})$ & -- & Optimized decoupling unitary. \\
        $\widetilde{U}(\boldsymbol{\theta_2^*})$ & -- & Optimized reduced unitary operator \\ & & that approximates the target model. \\
     \end{tabular}    
    
     \vspace*{0.5em}
     {\bf Steps:}
           
            \begin{enumerate}
                \item Randomly initialize the parameterized circuits for the decoupling operation $V(\boldsymbol{\theta_1})$ and the reduced update function $\widetilde{U}(\boldsymbol{\theta_2})$.
                \item Evaluate the combined cost function $C(\boldsymbol{\theta_1}, \boldsymbol{\theta_2})$ (see Eq.~\eqref{eq:combined_cost} below).
                \item Update parameters $\boldsymbol{\theta_1}$ and $\boldsymbol{\theta_2}$ toward the minimization of $C(\boldsymbol{\theta_1}, \boldsymbol{\theta_2})$ using gradient-based optimization (e.g., via the parameter-shift rule) or gradient-free methods.
                \item Repeat (b)-(d) until $C(\boldsymbol{\theta_1}, \boldsymbol{\theta_2})$ converges or meets a predefined threshold.
            \end{enumerate}
    \end{tcolorbox}
\end{figure}

\inlineheading{Quantum dimension reduction via variational quantum circuit Decoupling} Our variational training procedure involves two optimization procedures: (1) a decoupling transformation, $V(\boldsymbol{\theta}_1)$, intended to isolate the most relevant memory subspace, of dimension $\tilde{d}_q$, and expunge irrelevant degrees of freedom; and (2) an update unitary for the RQM, $\widetilde{U}(\boldsymbol{\theta}_2)$, acting solely within this compressed memory space, to approximate the  target observable dynamics.

Within an RQM, the significance of a specific memory state is proportional to its frequency of occurrence. Consequently, we prioritize reconstructing the action of $U$ over a weighted ensemble of memory states, rather than pursuing a global decoupling of the unitary operator that implicitly weighs all states equally~\cite{ximing2024}. 

To generate the ensemble of memory states $\{\ket{s_i}\}$, we adopt the following procedure. First, initialize the memory system to $\ket{0}$, and then iteratively apply the update unitary $U$ to the combined memory and output systems, measuring and resetting the output register between iterations. After a sufficient number of iterations, the memory system tends towards a pure state, $\ket{s_i}$, effectively encoding the memory state associated with the observed sequence as its history. By repeating this sampling procedure $K$ times, we obtain a finite ensemble of quantum memory states, $\{\ket{s_i}\}_{i=1}^K$. Since these states are sampled according to their frequency of occurrence during long-term evolution, they naturally form a representative weighted ensemble suitable for training the reduced model. Furthermore, in the asymptotic limit as $K \to \infty$, the ensemble average converges to the stationary density operator, $\rho_M$.

Given each sampled memory state $\ket{s_i}$, our training procedure is as follows. We first apply the parameterized decoupling unitary $V(\boldsymbol{\theta}_1)$, intended to partition the memory Hilbert space into a discarded subspace $\mathcal{H}_T$ and a retained memory subspace $\mathcal{H}_{\tilde{M}}$. This procedure is closely related to the mechanism of quantum autoencoders~\cite{romeroQuantumAutoencodersEfficient2017}, where a unitary learns to compress quantum states by mapping irrelevant degrees of freedom into a fixed fiducial state. The effectiveness of this decoupling is quantified by how closely the `trashed' subsystem in $\mathcal{H}_T$ is mapped to the fiducial state $\ket{0}_T$:
\begin{equation}
    D_i(\boldsymbol{\theta}_1) 
    = \bra{0}_T \operatorname{Tr}_{\tilde{M}}\!\left[ V(\boldsymbol{\theta}_1) \ketbra{s_i}{s_i} V(\boldsymbol{\theta}_1)^\dagger \right] \ket{0}_T.
\end{equation}

Subsequently, we train the reduced update unitary $\widetilde{U}(\boldsymbol{\theta}_2)$ to emulate the dynamics of the original model $U$ within the compressed subspace. Specifically, we evolve a combined system comprising the memory state $\ket{s_i}$ and the output register initialized to $\ket{0}$, under the original unitary $U$. We then apply the decoupling operator $V(\boldsymbol{\theta}_1)$ and trace out the discarded subsystem $T$, yielding the reduced state:
\begin{equation}
    \rho_i = \operatorname{Tr}_T\!\left[ V(\boldsymbol{\theta}_1) U \left(\ketbra{s_i}{s_i} \otimes \ketbra{0}{0}_{\text{out}}\right) U^\dagger V(\boldsymbol{\theta}_1)^\dagger \right].
\end{equation}
If $V(\boldsymbol{\theta}_1)$ effectively decouples the memory and $\widetilde{U}(\boldsymbol{\theta}_2)$ accurately captures the dynamics of $U$, then applying the inverse reduced unitary $\widetilde{U}(\boldsymbol{\theta}_2)^\dagger$ should revert the system to the initial compressed configuration (see Fig.~\ref{fig:cost}). We define this target reference state as:
\begin{equation}
    \sigma_i(\boldsymbol{\theta}_1) = \operatorname{Tr}_T\!\left[ V(\boldsymbol{\theta}_1) \ketbra{s_i}{s_i} V(\boldsymbol{\theta}_1)^\dagger \right] \otimes \ketbra{0}{0}_{\text{out}}.
\end{equation}
The performance of $\widetilde{U}$ is thus measured by the fidelity between the recovered state and the reference state:
\begin{equation}
    F_i(\boldsymbol{\theta}_1, \boldsymbol{\theta}_2)
    = \mathcal{F}\!\left( \widetilde{U}(\boldsymbol{\theta}_2)^\dagger \rho_i \, \widetilde{U}(\boldsymbol{\theta}_2), \, \sigma_i(\boldsymbol{\theta}_1) \right),
\end{equation}
where $\mathcal{F}(\rho,\sigma)$ denotes the cosine similarity metric defined in Eq.~\eqref{eq:cosine_similarity}.

To train the model, we construct a global cost function that aggregates performance over the sampled ensemble $\{\ket{s_i}\}_{i=1}^K$. Since our goal is to maximize both the decoupling effectiveness $D_i$ and the dynamical fidelity $F_i$, we define the cost function $C(\boldsymbol{\theta}_1, \boldsymbol{\theta}_2)$ as the negative weighted average of these quantities:
\begin{equation}\label{eq:combined_cost}
    C(\boldsymbol{\theta}_1, \boldsymbol{\theta}_2) 
    = -\frac{1}{K}\sum_{i=1}^K \Big[ \alpha\,D_i(\boldsymbol{\theta}_1) 
    + \beta\,F_i(\boldsymbol{\theta}_1,\boldsymbol{\theta}_2) \Big],
\end{equation}
where the hyperparameters $\alpha, \beta > 0$ govern the trade-off between the two competing objectives: (i) ensuring the trash subsystem $T$ is effectively decoupled by $V(\boldsymbol{\theta}_1)$, and (ii) ensuring the reduced unitary $\widetilde{U}(\boldsymbol{\theta}_2)$ accurately simulates the dynamics of $U$ within the retained memory subspace. Minimizing $C(\boldsymbol{\theta}_1, \boldsymbol{\theta}_2)$ therefore yields a model that is both dimensionally compressed and dynamically faithful. The parameters are updated using standard variational optimization strategies, such as gradient descent via the parameter-shift rule or gradient-free methods. A schematic overview of this training loop is presented in Fig.~\ref{fig:cost}.

\begin{figure}[htbp]
    \centering
    \includegraphics[width=0.95\columnwidth]{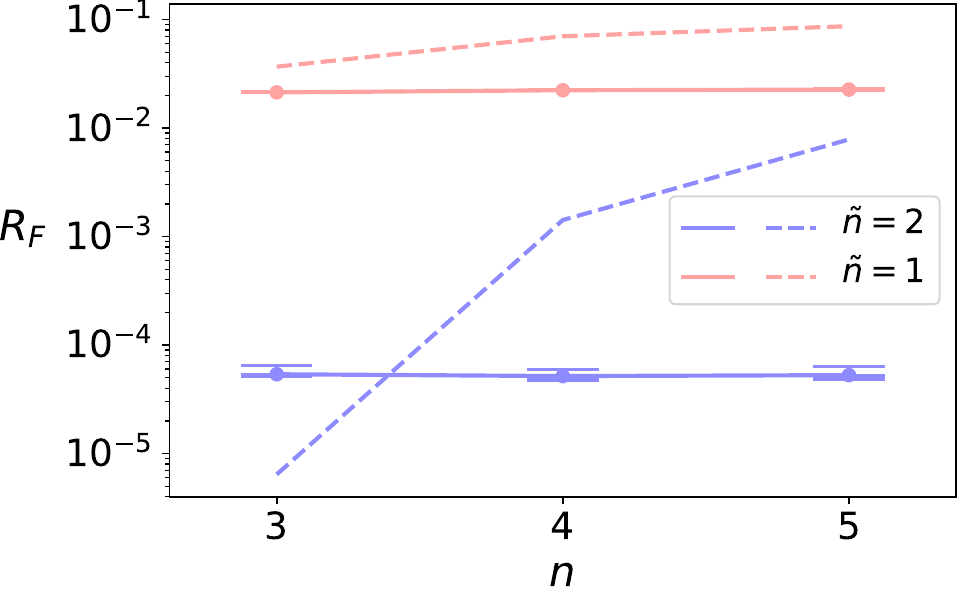}

    \caption{Quantum fidelity divergence rate \(R_F\) versus original memory size \(n\) for the cyclic walk model. Solid lines denote our approach, dashed lines the baseline; colors indicate reduced dimensions (\(\tilde{n}=2\) in blue, \(\tilde{n}=1\) in orange). Our method achieves consistently lower divergence, especially under stronger compression. Error bars show one standard deviation.
}
    \label{fig:cyclicwalk}
\end{figure}

\inlineheading{Example: cyclic walk model} To demonstrate the efficacy of our variational memory reduction framework, we apply it to quantum models simulating a discrete cyclic random walk on a ring of $N$ sites~\cite{garnerProvablyUnboundedMemory2017}. In the classical regime, the walker's position at time step $t$ is denoted by $X_t \in \{0, 1, \dots, N-1\}$, such that the full sequence $\mathcal{P} = \{X_t\}_{t \in \mathbb{Z}}$ characterizes the process statistics. The dynamics are governed by a transition matrix with elements $p_{kj} = P(X_{t+1} = k \mid X_t = j)$.


Previous studies indicate that quantum models require $n=\log N$ memory qubits to exactly simulate these stochastic processes~\cite{garnerProvablyUnboundedMemory2017}. However, the quantum statistical complexity, $C_q$, of these models is significantly lower than the number of qubits of the physical memory system. This discrepancy implies that the dynamics can be faithfully approximated within a smaller subspace. Treating the quantum model as a black box, we apply our variational truncation algorithm to compress the memory into a reduced register of $\tilde{n} \in \{1, 2\}$ qubits. We optimize the parameterized unitaries $V(\boldsymbol{\theta}_1)$ and $\widetilde{U}(\boldsymbol{\theta}_2)$ using the L-BFGS optimizer to minimize the unified cost function. The reconstruction quality is quantified using the quantum fidelity divergence rate, $R_F$ \eqref{eq:qfdr}, and we benchmark our framework against a baseline variational Matrix Product State (MPS) truncation approach~\cite{yangDimensionReductionQuantum2025}.

Figure~\ref{fig:cyclicwalk} compares the performance of our algorithm with the variational MPS truncation method by plotting the error metric $R_F$ as a function of $n$. Results are presented for compressed memory configurations $\tilde{n}=1$ and $\tilde{n}=2$. While the MPS truncation method (dashed lines) displays a clear upward trend—indicating that error grows significantly with $n$—our algorithm (solid lines) maintains a consistently low error rate throughout the observed range. This stability aligns with the observation that the $C_q$ of the original process remains stable as $n$ increases. Furthermore, whereas conventional MPS-based techniques typically necessitate either full access to the Matrix Product State representation or the preparation of long-range $L$-step entangled states for tomography, our method circumvents these requirements. By exploiting the intrinsic recurrent structure of the process, we avoid the generation of large-scale multipartite entanglement.


\inlineheading{Discussion} This work presents a variational quantum dimensionality reduction framework for recurrent quantum models (RQMs), designed to compress the internal memory structure of quantum sequential processes while preserving their dynamical behavior. The framework relies on two parameterized quantum circuits: a decoupling unitary $V(\boldsymbol{\theta}_1)$, which isolates the dynamically relevant memory subspace from redundant degrees of freedom, and a reduced recurrent unitary $\widetilde{U}(\boldsymbol{\theta}_2)$, which emulates the original process within this compressed space. These circuits are optimized via a unified cost function that balances decoupling efficiency with dynamical accuracy, as quantified by the quantum fidelity divergence rate (QFDR) - a metric tracking the accumulated distortion per time step~\cite{Yang2020, Bram2021, yangDimensionReductionQuantum2025}. Numerical simulations of a cyclic random walk~\cite{garnerProvablyUnboundedMemory2017} indicate that this approach achieves a reduction in QFDR of one to three orders of magnitude compared to conventional matrix product state (MPS) truncation~\cite{yangDimensionReductionQuantum2025}. Crucially, because the method operates solely on sampled trajectories, it circumvents the need for full access to the underlying tensor-network representation, offering a scalable and experimentally viable path for the dimension-reduced simulation of recurrent processes.

While prior research in quantum process compression and quantum statistical complexity has established that quantum models can represent stochastic processes with strictly lower memory requirements than their classical counterparts~\cite{Liu2019, elliott2020extreme,elliott2021quantum}, the present framework advances this paradigm through a distinct methodological approach. Rather than presupposing explicit knowledge of the full process dynamics or relying on analytic state reconstruction, we adopt a data-driven variational strategy that operates directly on the black-box unitaries generating the process. Furthermore, by avoiding the generation of long $q$-samples, our algorithm significantly reduces circuit depth and overall implementation complexity.

There are several promising avenues for further investigation. Meanwhile, extending the approach to include non-unitary dynamics or non-trivial input tapes would significantly broaden its applicability, enabling us to reduce memory costs for modeling non-Markovian quantum processes~\cite{Modi2018}, realize quantum reservoir computers~\cite{ghosh2019quantum}, data-sequence generation~\cite{vieira2022temporal}, and quantum agents executing complex adaptive games~\cite{elliottQuantumAdaptiveAgents2022}. Application of our procedures in settings where quantum recurrent models exhibit a scaling memory advantage~\cite{garnerProvablyUnboundedMemory2017,elliott2021quantum,elliottQuantumAdaptiveAgents2022,thmpson2025,aghamohammadiExtremeQuantumAdvantage2017} to see how well that advantage holds under shallow-circuit antsaz (e.g., ones in which the repeated unitary block is constant or logarithmic in classical complexity). From an experimental perspective, implementing the framework on noisy intermediate-scale quantum (NISQ) devices would provide a direct assessment of its robustness against decoherence, gate imperfections, and sampling noise. Such studies could also inform noise-aware modifications to the cost function or training protocol~\cite{hoRobustInferenceMemory2020,wangCanErrorMitigation2024,liuStochasticNoiseCan2025}. Finally, it would be exciting to apply such models to processes that exhibit uniquely quantum effects (e.g., contextuality, temporal Bell-inequalities~\cite{emary2014leggett}), providing a new toolkit to probe the underlying memory costs of causal structure.

\inlineheading{Acknowledgments} This work is supported by the National Research Foundation of Singapore through the NRF Investigatorship Program (Award No. NRF-NRFI09-0010), the National Quantum Office, hosted in A*STAR, under its Centre for Quantum Technologies Funding Initiative (S24Q2d0009), and the Singapore Ministry of Education Tier 1 Grant RG91/25 and RT4/23.  TJE is supported by the University of Manchester Dame Kathleen Ollerenshaw Fellowship. C.Y. is funded by Schmidt Sciences, LLC. C.L. is supported by the China Scholarship Council program (202306070092). Google Gemini was used to assist with text refinement in the preparation of this manuscript.


\onecolumngrid
\appendix
\newpage
{\hfill\centering \LARGE\textbf{Supplementary Material}\hfill}
\vspace{6em}

\section{Variational Truncation method}\label{app:baseline}

The variational method~\cite{Bram2021,yangDimensionReductionQuantum2025} iteratively optimizes the parameters of an MPS $\tilde{A}$ to maximize its overlap with a given reference MPS $A$:

\begin{equation}
\max_{\tilde{A}} \frac{|\langle \psi(\tilde{A}) | \psi(A) \rangle|^2}{\langle \psi(\tilde{A}) | \psi(\tilde{A}) \rangle}
\end{equation}

The algorithm involves 5 steps:

\begin{enumerate}
	\item Randomly initialize tensor $\tilde{A}$ with dimension $\tilde{d}$, compute the left and right Matrix Product State/Canonical Form of $\tilde{A}, \tilde{A}_l$ and $\tilde{A}_r$, and (2) the Matrix Product State/Mixed Gauge Form $\{ \tilde{A}_c, \tilde{C} \}$.

	\item Compute leading eigenvalue and eigenvectors $\eta, G_l, G_r$ according to
	\begin{equation}
	G_l \bar{E}_l = \eta G_l, \qquad \bar{E}_r G_r = \eta G_r
	\end{equation}
	where $\bar{E}_l$ and $\bar{E}_r$ are defined as:
	\begin{equation}
	\bar{E}_l = \sum_x A_l^x \otimes (\tilde{A}_l^x)^*, \qquad
	\bar{E}_r = \sum_x A_r^x \otimes (\tilde{A}_r^x)^*
	\end{equation}

	\item Update $\tilde{A}_c \to G_l \tilde{A}_c G_r$, $\tilde{C} \to G_l \tilde{C} G_r$.

	\item Evaluate $\Delta$, defined as:
	\begin{equation}
	\Delta := \left| \frac{\tilde{A}_c}{\eta} - \tilde{A}_l \tilde{C} \right|
	\end{equation}

	\item Repeat 2--5 until $\Delta < \Delta_{\text{thresh}}$.
\end{enumerate}

\section{Variational Circuit Ansatz}
\label{app:ansatz}

\begin{figure}[ht]
	\centering
	\includegraphics[width=0.5\columnwidth]{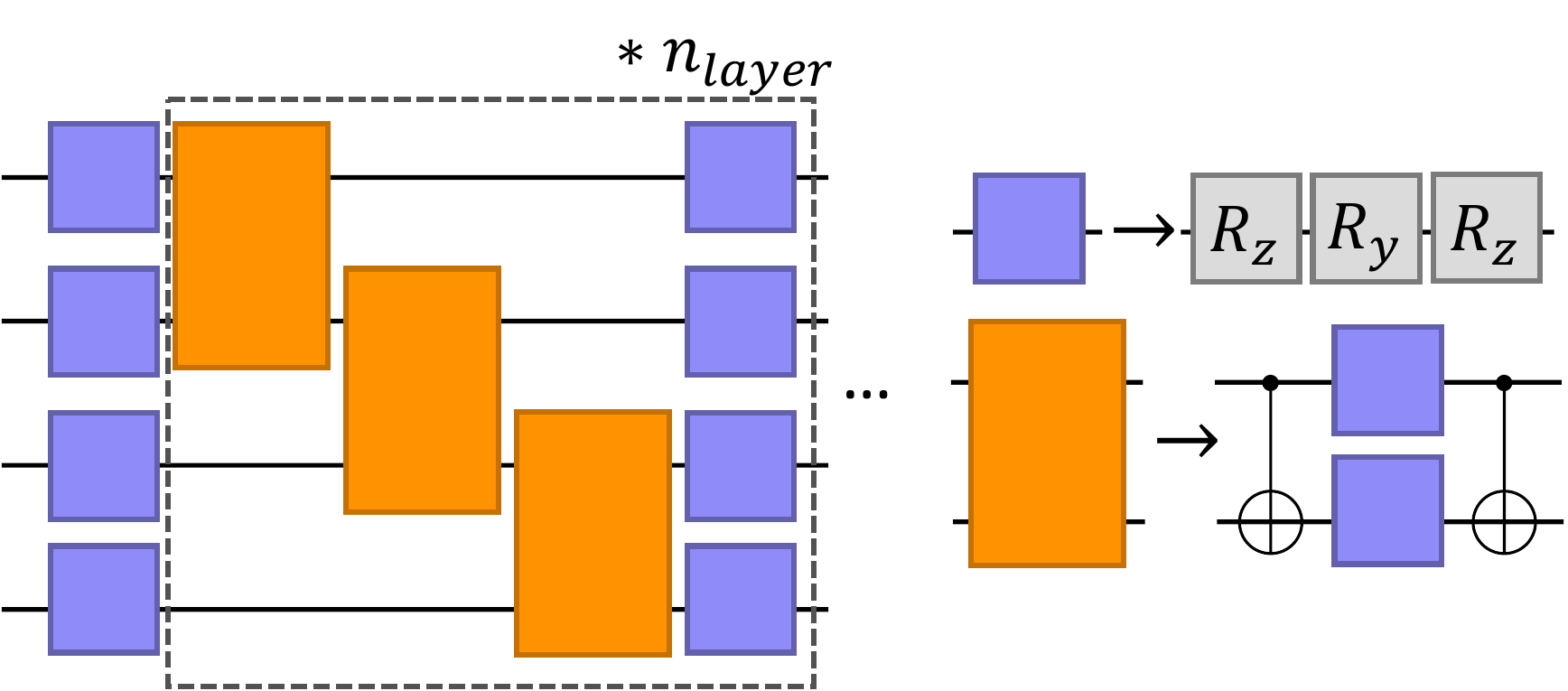}
	\caption[]{Variational ansatz used for both $V(\boldsymbol{\theta_1})$ and $\widetilde{U}(\boldsymbol{\theta_2})$. 
	Each layer consists of single-qubit $U3$ rotations followed by alternating two-qubit entangling blocks across adjacent qubits. 
	}
	\label{fig:ansatz}
\end{figure}

The parameterized quantum circuits used to represent the decoupling unitary $V(\boldsymbol{\theta_1})$ and the reduced update unitary $\widetilde{U}(\boldsymbol{\theta_2})$ share the same layered architecture, depicted in Fig.~\ref{fig:ansatz}. 
Each layer begins with single-qubit rotations on all qubits, followed by a chain of two-qubit entangling operations between nearest neighbors. 
The connectivity alternates between layers, ensuring full entanglement across the register after multiple repetitions.

The total number of trainable parameters in this ansatz is
\begin{equation}
	N_\text{param}(n_\text{qubit}, n_\text{layer}) 
	= n_\text{qubit}(3 + 9\,n_\text{layer}) - 6\,n_\text{layer},
\end{equation}
corresponding to $3n_\text{qubit}$ parameters for the initial rotation layer and $(9n_\text{qubit}-6)$ per additional layer.

This ansatz was chosen for its balance between expressivity and trainability. 
Both $V(\boldsymbol{\theta_1})$ and $\widetilde{U}(\boldsymbol{\theta_2})$ use this same circuit topology, differing only in the number of qubits and layers depending on the reduction target.

\section{Cyclic random walk model}\label{app:cyclic_walk}

\begin{figure}[ht]
	\centering
	\includegraphics[width=0.5\columnwidth]{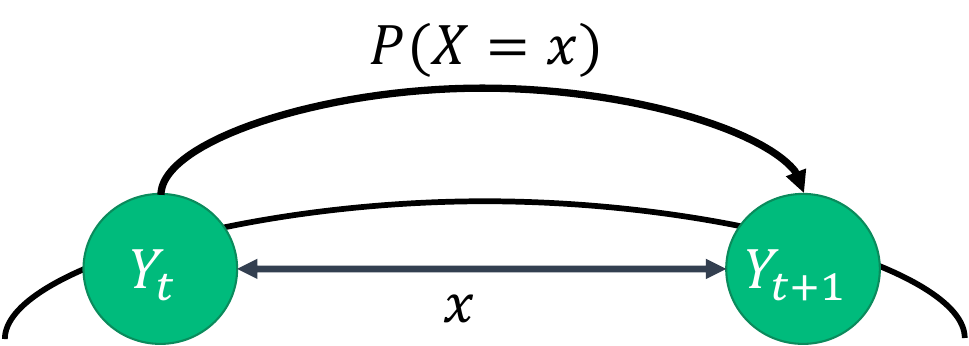}
	\caption{Schematic of the cyclic random walk on a ring of circumference one. 
	At each timestep, the walker at position \( Y_t \) shifts by a random increment \( X \) drawn from the distribution \( Q(X) \), wrapping around the ring as necessary.}
	\label{fig:cyclic_walk}
\end{figure}

We consider the cyclic random walk, a simple stationary stochastic process describing the motion of a particle confined to a ring of circumference one, as depicted in Fig.~\ref{fig:cyclic_walk}.  
At each discrete time \( t \in \mathbb{Z} \), the position of the particle is given by a random variable \( Y_t \in [0,1) \).  
Its evolution is driven by a real random variable \( X \), drawn independently at each timestep from a fixed probability density \( Q(X) \), such that
\begin{equation}
	Y_{t+1} = \mathrm{frac}(Y_t + X),
	\label{eq:continuous_cyclic}
\end{equation}
where \( \mathrm{frac}(y) = y - \lfloor y \rfloor \) denotes the fractional part of \( y \), so that positions differing by an integer represent the same physical location on the ring.  
We refer to \( Q(X) \) as the shift distribution and assume it has no explicit time dependence; the increments \( X \) are thus i.i.d. and independent of the current value of \( Y_t \), ensuring that the process is stationary. The dynamics of the continuous cyclic random walk can be viewed within the general framework of stochastic processes: the bi-infinite sequence \( \mathcal{P} = \{Y_t\}_{t \in \mathbb{Z}} \) is characterized by the joint distribution \( P(\overleftarrow{Y},\overrightarrow{Y}) \), where
\(\overleftarrow{Y} = \dots Y_{-2}Y_{-1}Y_0\) and
\(\overrightarrow{Y} = Y_1Y_2\dots\) denote past and future trajectories, respectively.  

To implement this process within a finite-dimensional recurrent quantum model, we restrict the position to a discrete set of \( N = 2^{n} \) equally spaced values
\begin{equation}
	y_j = \frac{j}{N}, \qquad j = 0, 1, \dots, N-1,
\end{equation}
each representing a site on the ring and corresponding to a distinct memory configuration encoded in a \( d_q \)-qubit register.  
At each timestep, the walker at site \( y_j \) transitions to site \( y_k \) with probability
\begin{equation}
	p_{kj} = P\!\bigl(Y_{t+1} \in \mathcal{I}_k \,\big|\, Y_t = y_j\bigr),
	\label{eq:transition_prob_app}
\end{equation}
where the interval
\begin{equation}
	\mathcal{I}_k = \Bigl\{ y \in [0,1) : \bigl|y - y_k\bigr| < \frac{1}{2N} \Bigr\}
\end{equation}
collects all continuous positions that are ``rounded'' to the discrete site \( y_k \).  
Given Eq.~\eqref{eq:continuous_cyclic}, these transition probabilities are obtained by integrating the shift distribution \( Q(X) \) over all shifts that move a point in \( \mathcal{I}_j \) into \( \mathcal{I}_k \); index arithmetic is taken modulo \( N \) to respect the ring structure.  

The resulting discretized cyclic walk defines a finite-state Markov process, as the conditional distribution of the next position depends only on the current site:
\begin{equation}
	P(Y_{t+1} \mid Y_t, Y_{t-1}, \dots) = P(Y_{t+1} \mid Y_t).
\end{equation}
Letting \( \mathcal{P}_N \) denote the corresponding discrete process for a given \( N \), as the resolution is increased (\( N \to \infty \), equivalently \( n \to \infty \)), the family \( \{\mathcal{P}_N\} \) converges in distribution to the continuous dynamics of Eq.~\eqref{eq:continuous_cyclic}.
This construction provides a natural bridge between the continuous cyclic random walk and the finite-state model used in the main text; it is this discrete version (with \( N = 2^n\) sites) that we embed into our recurrent quantum models.

\bibliographystyle{unsrt}
\bibliography{refs}

\end{document}